\documentclass[sn-mathphys]{sn-jnl}

\jyear{2022}%

\theoremstyle{thmstyleone}%
\theoremstyle{thmstyletwo}%

\theoremstyle{thmstylethree}%

\usepackage[normalem]{ulem}
\usepackage{soul}
\definecolor{antiquefuchsia}{rgb}{0.57, 0.36, 0.51}
\definecolor{burgundy}{rgb}{0.5, 0.0, 0.13}

\definecolor{MyDarkGreen}{rgb}{0.02,0.60,0.06}

\begin{document}

\title[Potts Model with Invisible States: A Review]{Potts Model with Invisible States: A Review}

\author*[1,2,3]{\fnm{Mariana} \sur{Krasnytska}}\email{kras.marjana@gmail.com}

\author[2,3]{\fnm{Petro} \sur{Sarkanych}}\email{petrosak@gmail.com}

\author[1,3]{\fnm{Bertrand} \sur{Berche}}\email{bertrand.berche@univ-lorraine.fr}

\author[2,3,4]{\fnm{Yurij} \sur{Holovatch}}\email{hol@icmp.lviv.ua}

\author[2,4]{\fnm{Ralph} \sur{Kenna}}\email{r.kenna@coventry.ac.uk}

\affil*[1]{\orgdiv{Laboratoire de Physique et Chimie Th\'eoriques}, \orgname{Universit\'e de Lorraine - CNRS, UMR 7019}, \orgaddress{\city{Nancy},   \country{France}}}

\affil[2]{\orgname{Institute for Condensed Matter  Physics, National Acad. Sci. of Ukraine}, \orgaddress{\city{Lviv}, \postcode{79011}, \country{Ukraine}}}

\affil[3]{\orgname{$\mathbb{L}^4$ Collaboration \& Doctoral College for the Statistical Physics of Complex Systems}, \orgaddress{\city{Leipzig-Lorraine-Lviv-Coventry}, \country{Europe}}}

\affil[4]{\orgdiv{Centre for Fluid and Complex Systems}, \orgname{Coventry University}, \orgaddress{\city{Coventry}, \postcode{CV1 5FB}, \country{United Kingdom}}}


\abstract{The Potts model with invisible states was  introduced to explain discrepancies between theoretical predictions and experimental observations of phase transitions in some systems where  $Z_q$  symmetry is spontaneously broken. It differs from the ordinary $q$-state Potts model in that each spin, besides the  usual $q$ visible states, can be also in any of $r$  so-called invisible states. Spins in an invisible state do not interact with their neighbours but they do contribute to  the entropy  of the system. 
As a consequence,  an increase in $r$  may cause a phase transition to change from second to first order.  Potts models  with invisible states describe  a number of systems of interest in physics and beyond  and have  been  treated by various tools of statistical and mathematical physics. 
In this paper we aim to give a review of this fundamental topic.}


\maketitle

\section{Introduction}\label{I}

The Potts model with invisible states was introduced  in Refs.~\cite{Tamura2010,Tanaka2011} to explain discrepancies between theoretical predictions and experimental observations of peculiarities in the critical behaviour of certain 2D systems with spontaneously broken $Z_3$-symmetry.  
An example is given by ${\rm Bi_3Mn_4O_{12}(NO_3)}$ with magnetic elements located on a honeycomb lattice with
an underlying $Z_3$-symmetry, where a first order phase transition is observed experimentally \cite{Tamura08,Stoudenmire09,Okumura10}. 
This  is obviously incompatible with theoretical predictions for the standard 2D ferromagnetic $q$-state Potts model 
which, while for $q > 4$ has a first-order phase transition,  undergoes a second order  transition at finite temperature when 
$q \leq 4$ \cite{Wu82}.
 
 Within scarcely a decade of its appearance, the Potts model with invisible states attracted significant interest.
It was analysed by both analytic \cite{Enter11,Enter12,Johnston13,Mori12,Krasnytska16,Sarkanych17,Sarkanych18,Sarkanych19} and numerical \cite{Tanaka2011,Tamura2010,Tanaka11a} tools to explain diverse phenomena in physics and beyond \cite{Iakobelli12,Kiraly19,Lee17}. 
To a large extent, this interest is due to the prominent feature of the model that allows changing the strength and even the order of the phase transition by continuously tuning  the model parameters, keeping other global features like space dimensionality, interaction range and symmetry unchanged. 
	
The Hamiltonian of the invisible-states Potts model reads:
\begin{equation}\label{1.1}
H=-\sum_{<i,j>}\delta _{S_i,S_j}\sum_{\alpha=1}^q \delta
_{S_i,\alpha} {\delta _{S_j,\alpha}},
\end{equation}
where $S_i=1,\ldots,q,q+1,\ldots,q+r$ is the Potts variable, $\delta$ is the Kronecker delta, $q$ and $r$ are the numbers of visible and invisible states respectively. 
In the text, we will refer to this model as $(q,r)$ ISPM (the acronym standing for Invisible-State Potts Model). 
The first sum in Eq.~(\ref{1.1}) is taken over all distinct pairs of interacting particles, and the second sum requires both of the interacting spins to be in a state among the subset $S_i \in (1,\ldots,q)$.  
This is the reason to call such states the `visible' ones.
As one can see from the Hamiltonian (\ref{1.1}), the main and only difference from the ordinary Potts model \cite{Potts1952,Wu82} is that when a spin is in one of the $q+1,\ldots,q+r$ `invisible' states, it does not interact with its neighbours, and as a result does not contribute to the interaction energy.
However, their presence increases the number of possible spin configurations and hence they do contribute to the entropy. 
Obviously, these are the changes in parameter $r$ that govern the number of invisible states and in this way lead to the changes in entropy and ultimately cause new unusual behaviour of the model.
It should be noted that the number of ground states of the $(q,r)$ ISPM is the same as that of the standard ferromagnetic $q$-state Potts model. 
Therefore, the $(q,r)$ ISPM exhibits a phase transition with $q$-fold symmetry breaking.

Since its appearance, the ISPM  was used to describe a number of systems in physics and beyond. 
It has been  treated by various tools of statistical and mathematical physics. 
In this paper, we aim to give a review of the advances related to this model. 
To a certain extent, this review reflects our personal experience, therefore certain results in the field might be overlooked. 
This is by no means to suggest we underestimate their importance. 

The rest of the paper is organized as follows. 
In Section \ref{II} we present a review of the original papers where the model was proposed \cite{Tamura2010,Tanaka2011} and 
some other papers they inspired, discussing also rigorous results and MC simulations. Exact solution for the 1D ISPM 
and partition function zeros analysis are considered in Section \ref{III}. The subject of section
\ref{IV} is the ISPM on a complete graph and on a scale-free network. We end by the conclusions and outlook
in section \ref{V}.

\section{ISPM: formulation, rigorous results and MC simulations} \label{II}

 \begin{figure}[t]
  \begin{center}
   \includegraphics[width=0.9\columnwidth]{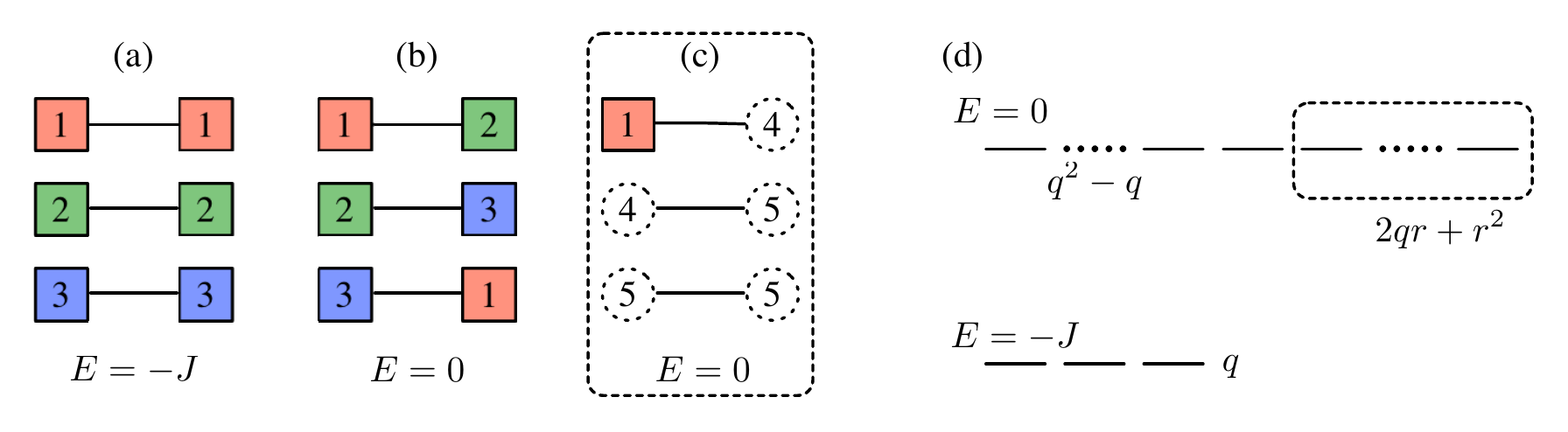}%
  \end{center}
  \caption{\label{fig:energy_twospins}
  (a)-(c): Different spin configurations of the ($3$,$2$) ISPM.
  The squares and dotted circles indicate visible (colored) states and invisible states, respectively.
  (a) Ground states of two spin system.
  (b) and (c) Excited states of two spin system.
  Both (a) and (b) also appear in the case of the standard ferromagnetic $q$-state Potts model 
  while configurations shown in (c) appear only due to invisible states.
  (d) Energy level structure of two spin system for the ($3$,$2$) ISPM.
  The number of ground states and excited states are $q$ and $q^2-q+2qr+r^2$, respectively.
  The dotted boxes in (c) and (d) denote the contribution of invisible states \cite{Tanaka2011}.
  }
 \end{figure}

Different models have been proposed to explain how the order of a phase transition can be changed by controlling various parameters. 
For example, the Blume-Capel \cite{Capel66,Blume66} and Blume-Emery-Griffiths  \cite{Blume71} models achieve this by changing the 
crystal field and the chemical potential. 
The link between the Blume-Emery-Griffiths model and $(2,r)$ ISPM was shown in Ref. \cite{Tamura2010}. 
In the Wajnflasz model \cite{Wajnflasz71} the number of microstates defines the order of the phase transition. 
However, none of these cases goes beyond the $Z_2$ symmetry. 


The ISPM was formulated \cite{Tamura2010,Tanaka2011} 
as an attempt to tune the phase transition strength/order in the case of $Z_q$ symmetry. 
A simple illustration of the energy levels of a two spin system with invisible states is given in Fig. \ref{fig:energy_twospins} (reproduced from Ref.~\cite{Tanaka2011}).
Figure \ref{fig:energy_twospins} (a) shows three different combinations of the ground state for the ($3$,$2$) ISPM.
Figure \ref{fig:energy_twospins} (b) and (c) denote excited states for the ($3$,$2$) ISPM.
States depicted in Fig.~\ref{fig:energy_twospins} (a) and (b) appear in the standard ferromagnetic $q$-state Potts model, whereas states depicted in Fig.~\ref{fig:energy_twospins} (c) appear only when the invisible states are introduced.
Figure \ref{fig:energy_twospins} (d) shows the energy level structure of the two spin system for the ($q$,$r$) ISPM.
The number of ground states is $q$ and that of the excited states is $q^2 - q + 2qr + r^2$.
The contribution of  invisible states in the excited states is $2qr + r^2$.
This energy level structure indicates that $q$-fold symmetry breaks at the transition point.

\begin{figure}[t]
  \begin{center}
   \includegraphics[width=0.5\columnwidth]{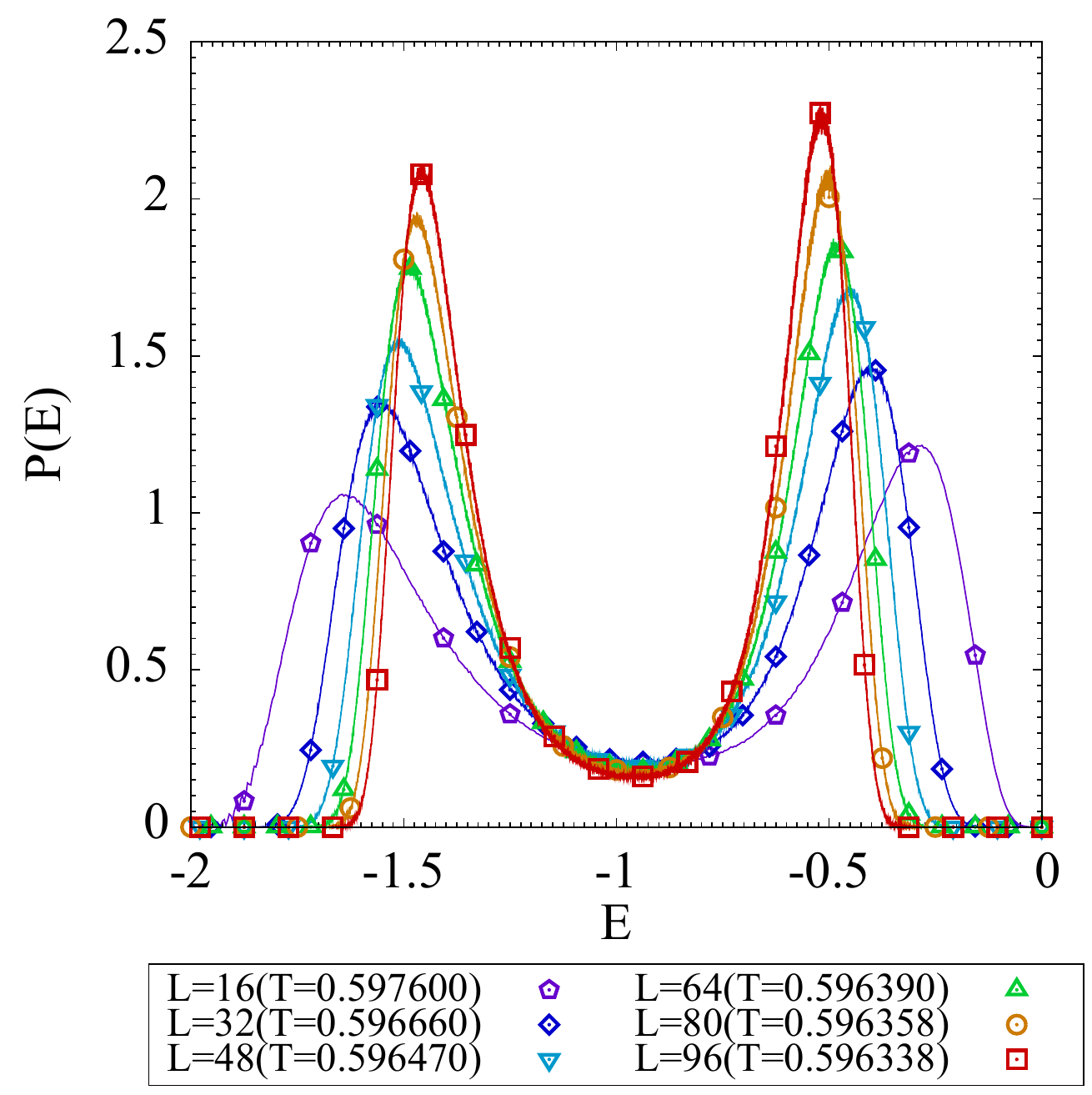}%
  \end{center}
  \caption{
  \label{graph:eng_histogram}
  (color online)
  Energy distribution for the simulations of ($3$,$25$) ISPM on a square lattice.  The double peak structure, where both peaks increase with the system size, signals the existence of a first order phase transition.
  The error bars are omitted for clarity since they are smaller than the symbol size \cite{Tamura2010}.
  } \label{fig3}
 \end{figure}

The Hamiltonian of the ISPM is given by Eq. \ref{1.1} and at $r=0$ reduces to the standard Potts model \cite{Potts1952}. 
In Ref. \cite{Tamura2010} the phase transition of the ferromagnetic $(q,r)$ ISPM on a square lattice with periodic boundary conditions was investigated by the Bragg-Williams approximation for  $q = 2, 3, 4$, when a second order phase transition is predicted by the theory. 
The authors also investigated the nature of the phase transition of the ($3,25$) ISPM by Monte Carlo simulations using the standard Metropolis method. 
They calculated the transition temperature and analysed  the specific heat, magnetization and  density of states.
Fig. \ref{fig3} shows the probability distribution of the internal energy $P(E)$ with the bimodal distribution form, which signals about the typical behavior of a first-order phase transition \cite{Tamura2010}.

\begin{figure}[h!]
  \begin{center}
   \includegraphics[scale=0.4]{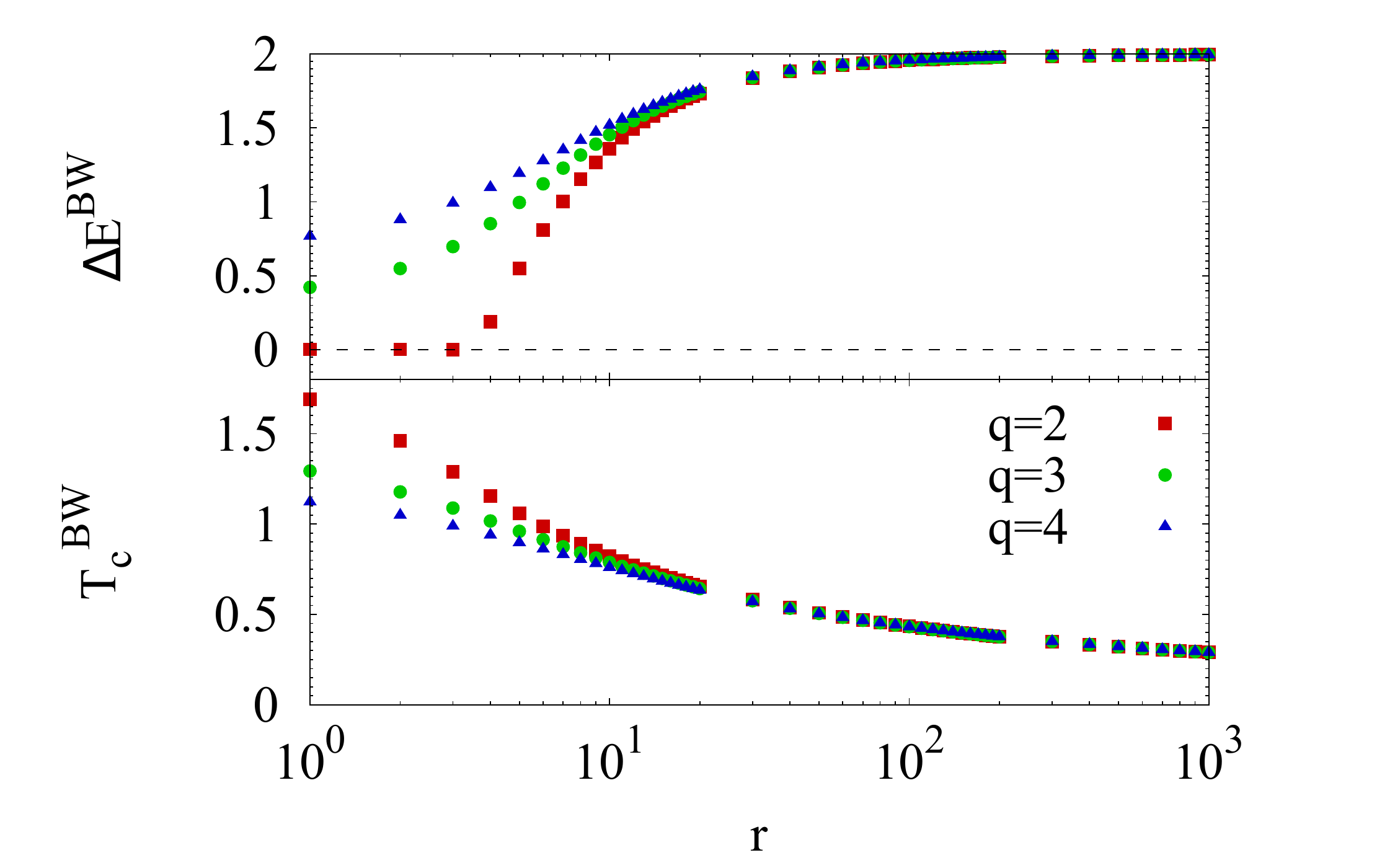}%
  \end{center}
 \caption{\label{fig:mf_tc_de}
  The transition temperature $T_{\rm c}^{\rm BW}$ and the latent heat $\Delta E^{\rm BW}$ as functions of the number of invisible states $r$ for $z=4$ obtained by the Bragg-Williams approximation \cite{Tanaka2011}.
}
 \end{figure}

In the subsequent paper \cite{Tanaka2011}, using the Bragg-Williams variant of the mean field  approximation and MC simulations, it was found that the transition temperature decreases and the latent heat increases as the number of invisible states increases. 
The dependence of the latent heat and of the transition temperature on the number of invisible states $r$ for the ISPM with $q = 2, 3$ and $4$ is shown in Fig. \ref{fig:mf_tc_de} reproduced from Ref.~\cite{Tanaka2011}). 
For this plot, it was assumed that each spin interacts with  $z = 4$ nearest neighbours.
From Fig. \ref{fig:mf_tc_de}, it follows that a second-order phase transition takes place only for $(2,1)$, $(2,2)$ and $(2,3)$ ISPM. 
The simulation results for the specific heat and transition temperature obtained in in Ref. \cite{Tanaka2011} are summarized in Table \ref{table:tc_lh}.

\begin{table}
\begin{center}
 \caption{\label{table:tc_lh}
 Transition temperature and latent heat for several parameter sets ($q$,$r$) \cite{Tanaka2011}.
 For $r=0$ case, $T_{\rm c}$ and $\Delta E$ were obtained exactly in Ref. \cite{Wu82}.}
 \begin{tabular}{c|ccc}
  \hline\hline
  ($q$,$r$) & $T_{\rm c}$ & $\Delta E$ & Symmetry \\
  \hline
  ($2$,$0$) & $1.13459$ & $0$ & 2-fold \\
  ($2$,$30$) & $0.57837(1)$ & $1.02(2)$ & 2-fold \\
  ($2$,$32$) & $0.56857(1)$ & $1.23(2)$ & 2-fold \\
  ($3$,$0$) & $0.994973$ & $0$ & 3-fold \\
  ($3$,$25$) & $0.59630(1)$ & $0.81(2)$ & 3-fold \\
  ($3$,$27$) & $0.58513(1)$ & $1.05(2)$ & 3-fold \\
  ($4$,$0$) & $0.910239$ & $0$ & 4-fold \\
   ($4$,$20$) & $0.61683(1)$ & $0.68(2)$ & 4-fold \\
  ($4$,$22$) & $0.60396(1)$ & $0.87(2)$ & 4-fold \\
  \hline\hline
 \end{tabular} 
\end{center}
\end{table}

A random-cluster representation of the ISPM was introduced in Refs. \cite{Enter11,Enter12,Iakobelli12}. 
It allows the authors using the Pirogov-Sinai argument to rigorously prove that when $(q+r)$ is large enough the system undergoes a first order phase transition. 
This can be achieved either through a large number of visible states $q$, or at small values of $q$ for a large number of invisible states $r$.  
It is possible to rewrite the partition function for the $(q,r)$ ISPM in terms of the partition function of a $r$-biased random-cluster model.
Just as the standard random-cluster model, the $r$-biased model is a correlated bond-percolation model. 
The difference between this new model and the original random-cluster model is that its singleton connected components here have weights $q+r$ while non-singleton connected components have weight $q$. 
In the limit $r\to0$ the original random cluster representation of the Potts model is recovered.
Larger weights of singleton connected components lead to an increase in entropy when  the number of invisible states increases. However, the ground state degeneracy (i.e., the number of visible states) remains constant with this change.  
The random-cluster representation of the Potts model allows for an elegant formulation of the intuitive concepts of “order” and “disorder” \cite{Laanait91}.
Each configuration in this model is treated as a certain arrangement of "ordered" and "disordered" bonds of a graph the model is considered in Ref.~\cite{Iakobelli12}.  
The equilibrium state is then defined by the lower energy caused by the interaction between ordered and disordered clusters and a first order phase transition occurs if there is a latent heat at the transition point.  
However, this approach does not allow for a precise finding of the  value $r_c$ at which the first-order transition occurs.
The transition temperature of the ISPM was asymptotically given by $\beta_c \sim \frac{1}{2} \log(q+r)$.  
The proofs were applied for quite high values of $q$ and/or $r$, which forced to assume that the dynamical properties of the ISPM have similar behaviours to what occurs in the ordinary Potts model \cite{Borgs99}.

\begin{figure}[t!]
\centerline{\includegraphics[width=0.6\columnwidth]{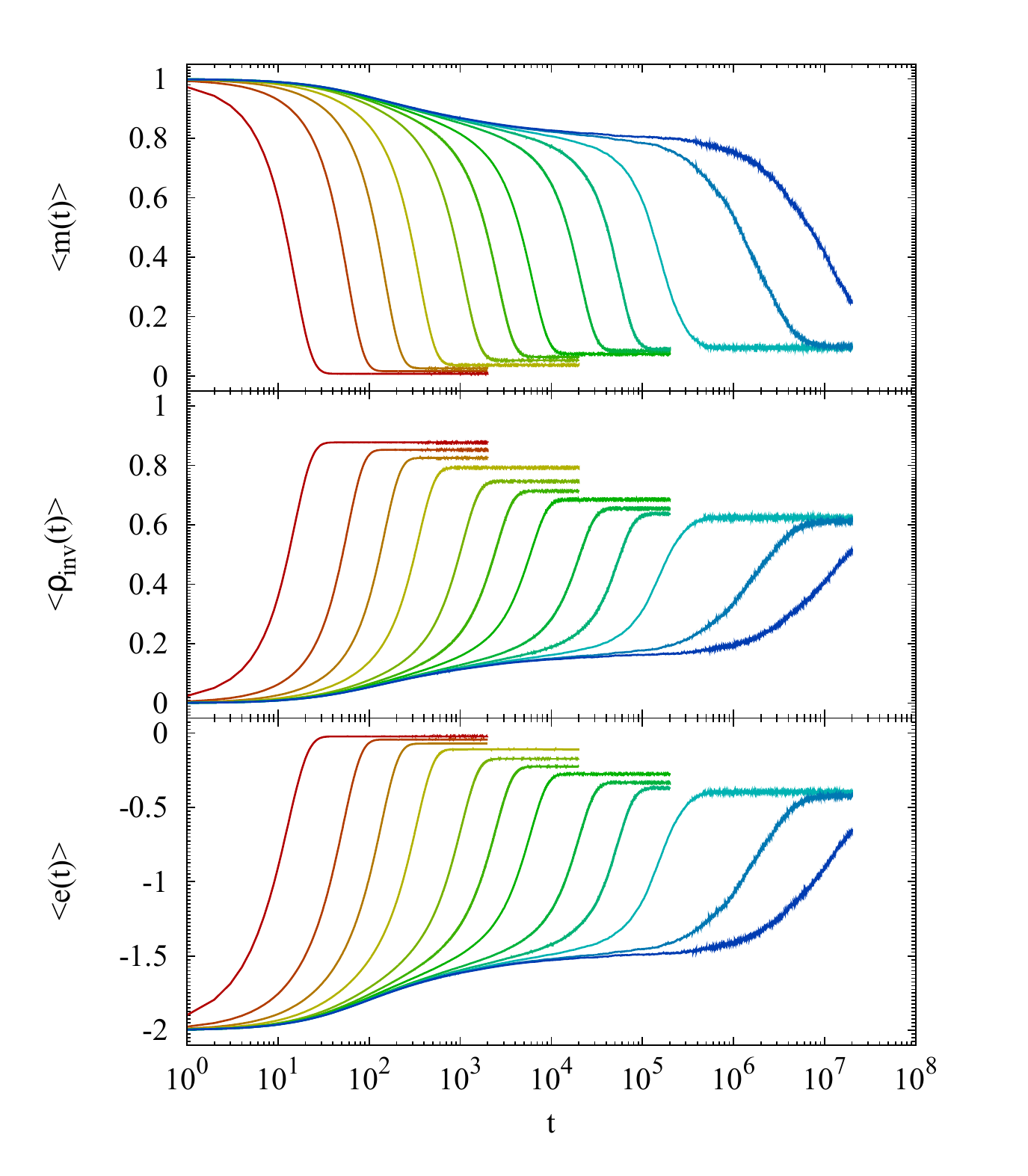}}
    \caption{Time evolution of the order parameter $\langle m(t)\rangle$, density of invisible states $\langle \rho_{inv}(t)\rangle$, 
    and the internal energy $\langle e(t)\rangle$ ($\langle ...\rangle$ of the  (3,27) ISPM 
    starting from the completely ordered state  for different fixed  temperatures.  \cite{Tanaka11a}}
    \label{Tanaka11adynamics}
\end{figure}

Melting dynamical properties of the ISPM on a square lattice were studied at fixed temperatures above the transition ones \cite{Tanaka11a}. 
The single-spin-flip Metropolis Monte Carlo method  was used for the time-evolution rule.
In order to analyse the dynamical properties of the model, three observables were examined -- the order parameter $\langle m(t)\rangle$, the density of invisible states $\langle \rho_{\rm inv}(t)\rangle$ and the internal energy $\langle e(t)\rangle$ \cite{Tanaka11a}. 
Here the $\langle ...\rangle$ means an ensemble average.
In Fig.\ref{Tanaka11adynamics} these three values for the (3,27) ISPM are plotted vs time for different temperatures. 
It is clear that there is a two-step relaxation,  typical in systems with a first-order phase transition and also in glassy systems.
Another interesting aspect is the behaviour of the order parameter. As temperature approaches the transition temperature, the plateau gets longer.

The characteristic melting time $\tau_{\rm max}$ was found from the peaks in dynamic susceptibility. 
As the temperature approaches the transition temperature the melting time increases. 
The growth speed of $\tau_{\rm max}$  depends on the energy barrier between the ordered  and paramagnetic states.
In the ISPM it is possible to control the energy barrier between the phases without changing the symmetry which breaks at the transition point. To investigate how invisible states affect the melting time, a $(3,25)$ ISPM was simulated. 
The results obtained indicate that far from the critical point there is no effect on the melting time, however near  $T_c$, the invisible states have to be taken into account as melting time increases with $r$.

\begin{figure} 
\centerline{\includegraphics[width=0.5\columnwidth]{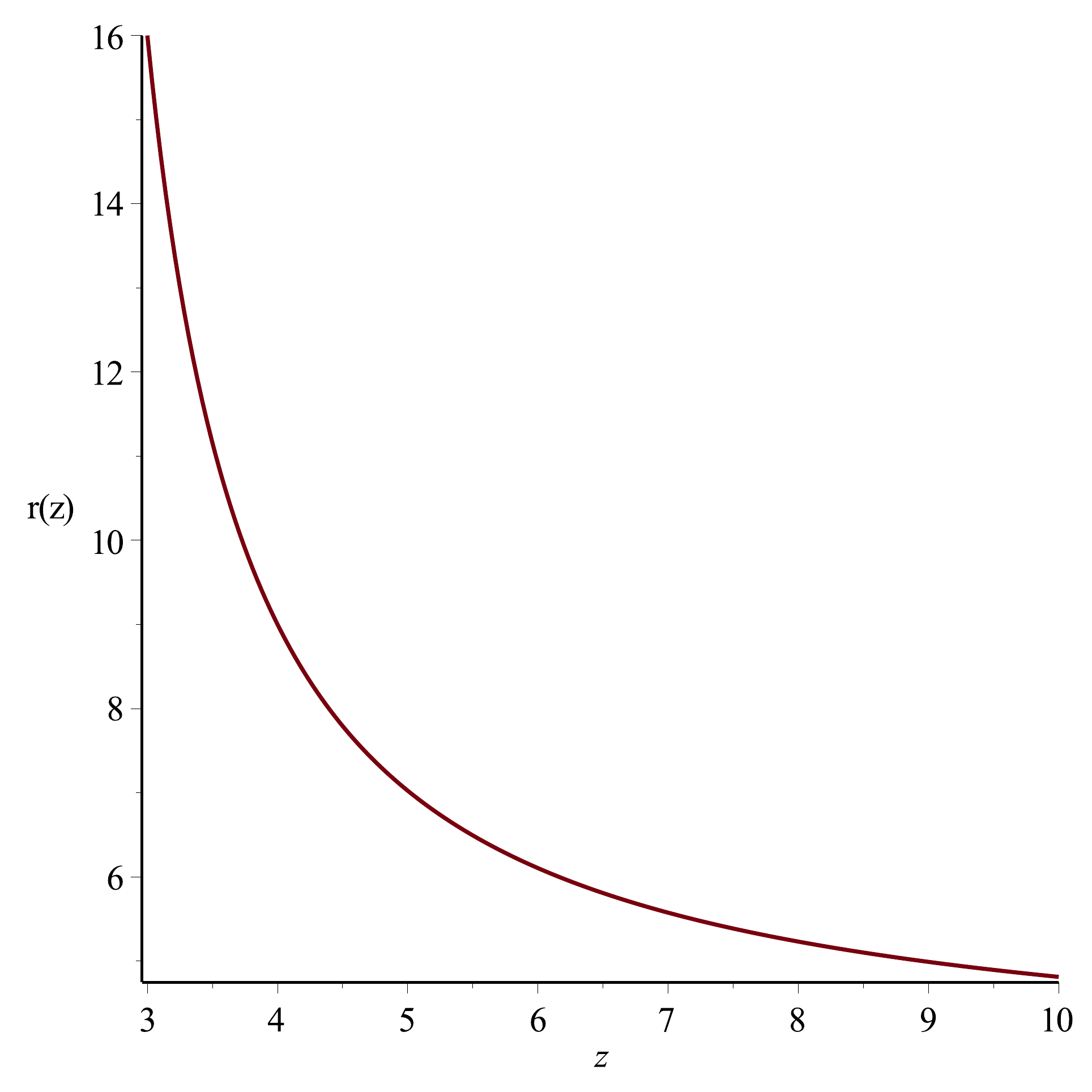}}
    \caption{The critical number of invisible states $r_c(z)$ above which the ($2,r$) ISPM displays a first order transition, plotted against the number of neighbours $z$ on the Bethe lattice \cite{Ananikian13}.}
    \label{ananikian_rc}
\end{figure}

In Ref. \cite{Ananikian13} ISPM was considered on a Bethe lattice with $z$ neighbours in a particular case $q=2$. 
It was shown that the marginal value of $r_c$ discriminating between the second and the first order transition regimes depends on the coordination number $z$. 
Using the aforementioned link with the Blume-Emery-Griffiths model the analytic expression for $r_c$ was obtained:
\begin{equation}
r_c(z) =  \cfrac{4 z}{3(z-1)} \left( \cfrac{z-1}{z-2}\right)^z\, .
\end{equation} 
For the particular values  $z=3$ and $z=4$, this gives $r_c=16$ and $r_c=9$, respectively. Whereas in the limit $z\to \infty$ 
one gets the Bragg-Williams, mean-field result $r_c=4e/3 \simeq 3.624$. The dependency of $r_c$ on $z$ is shown
in Fig. \ref{ananikian_rc}.  

In Ref. \cite{Mori12} the ISPM was used to illustrate the exactness of the mean-field theory in long-range interacting systems. 
For such classical spin systems with non-additive long-range interactions, it is expected that the entropy is identical to that of the corresponding mean-field model. 
This correspondence is sometimes called `exactness of the mean-field theory'. 
However, it was found that this statement is not necessarily true if the microcanonical ensemble is not equivalent to the canonical ensemble in the mean-field model. 
In Ref. \cite{Mori12} the necessary and sufficient conditions for exactness of the mean-field theory were  analysed in the case of the ISPM. 

An alternative mean-field approach was proposed for ISPM on thin graphs \cite{Johnston13}.
Employing 3-regular random (`thin') graphs the authors considered the Ising limit $q=2$ of the ISPM. 
The results showed that adding the invisible states changes the   transition from  second to  first order. 
However, the number of invisible states required to effect the transmutation for $q=2$ is $r_c=17$, which is much higher than in the Bragg-Williams case \cite{Tamura2010} or the value obtained for the limit $z\to\infty$ on the Bethe lattice~\cite{Ananikian13}.

\section{Exact solution for a linear chain and partition function zeros analysis}\label{III} 

Thermodynamics of any system can be recovered from its partition function.
Alternatively, to analyse its critical properties one can analyse the behaviour of the partition function zeros. 
The exact solution for the one-dimensional ISPM  was obtained 
in Refs.  \cite{Sarkanych17,Sarkanych18} using  transfer matrix methods. This section
briefly summarizes the properties of the solution as well as the
 behaviour of the partition function zeros 
in the complex plane in the presence of invisible states. 

For the 1D chain of $N$ spins  with nearest-neighbour interactions and periodic boundary conditions 
it is convenient to rewrite the Hamiltonian (\ref{1.1}) of the ISPM as:
\begin{equation}
H = -\sum_{i}\sum_{\alpha=1}^q\delta_{s_i,\alpha}\delta_{\alpha,s_{i+1}}   - h_1\sum_i \delta_{s_i,1} - h_2\sum_i\delta_{s_i,q+1} \, ,
\label{4}
\end{equation}
where the sum over $i$ is taken over all sites of the chain, and one has introduced two separate magnetic fields $h_1$ and $h_2$
 acting on the first visible state $s_i=1$,  and the first invisible state $s_i=q+1$  correspondingly. 
The transfer matrix formalism  is adopted to obtain an exact solution of the model (\ref{4}). 
To this end, the Hamiltonian (\ref{4}) is presented as a sum of terms representing one bond each, $H =\sum_{i} H_{i}$. 
Then, the partition function reads:
\begin{equation}\label{6}
Z=\sum_{s}\prod_{i}\exp\left( -\beta (\sum_{\alpha=1}^q\delta_{s_i,\alpha}\delta_{\alpha,s_{i+1}}+h_1\delta_{s_i,1}+h_2\delta_{s_i,q+1})\right)\, .
\end{equation}

\begin{figure}[h!]
	\begin{center}
		\includegraphics[width=0.35\columnwidth]{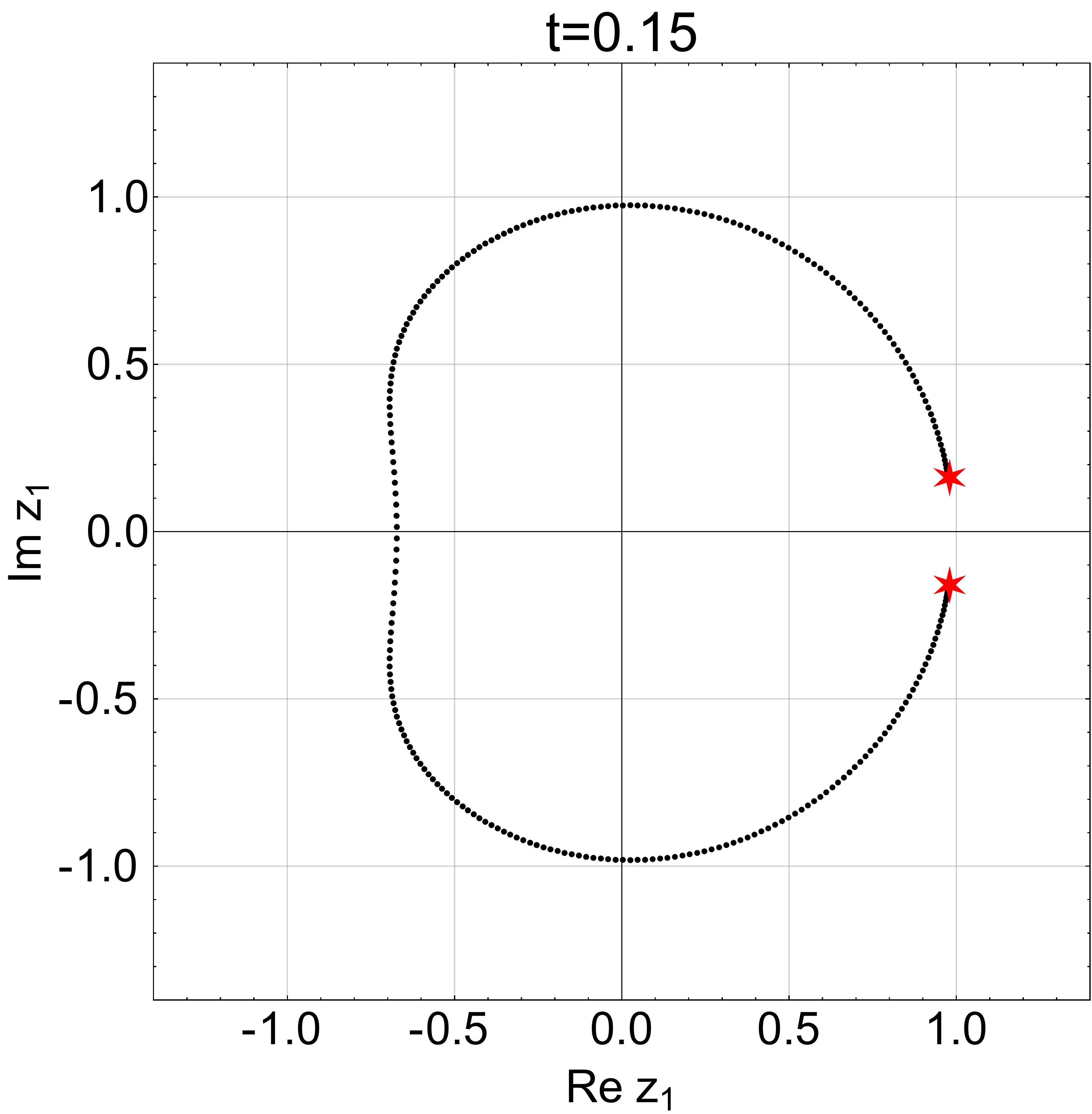}
		\includegraphics[width=0.35\columnwidth]{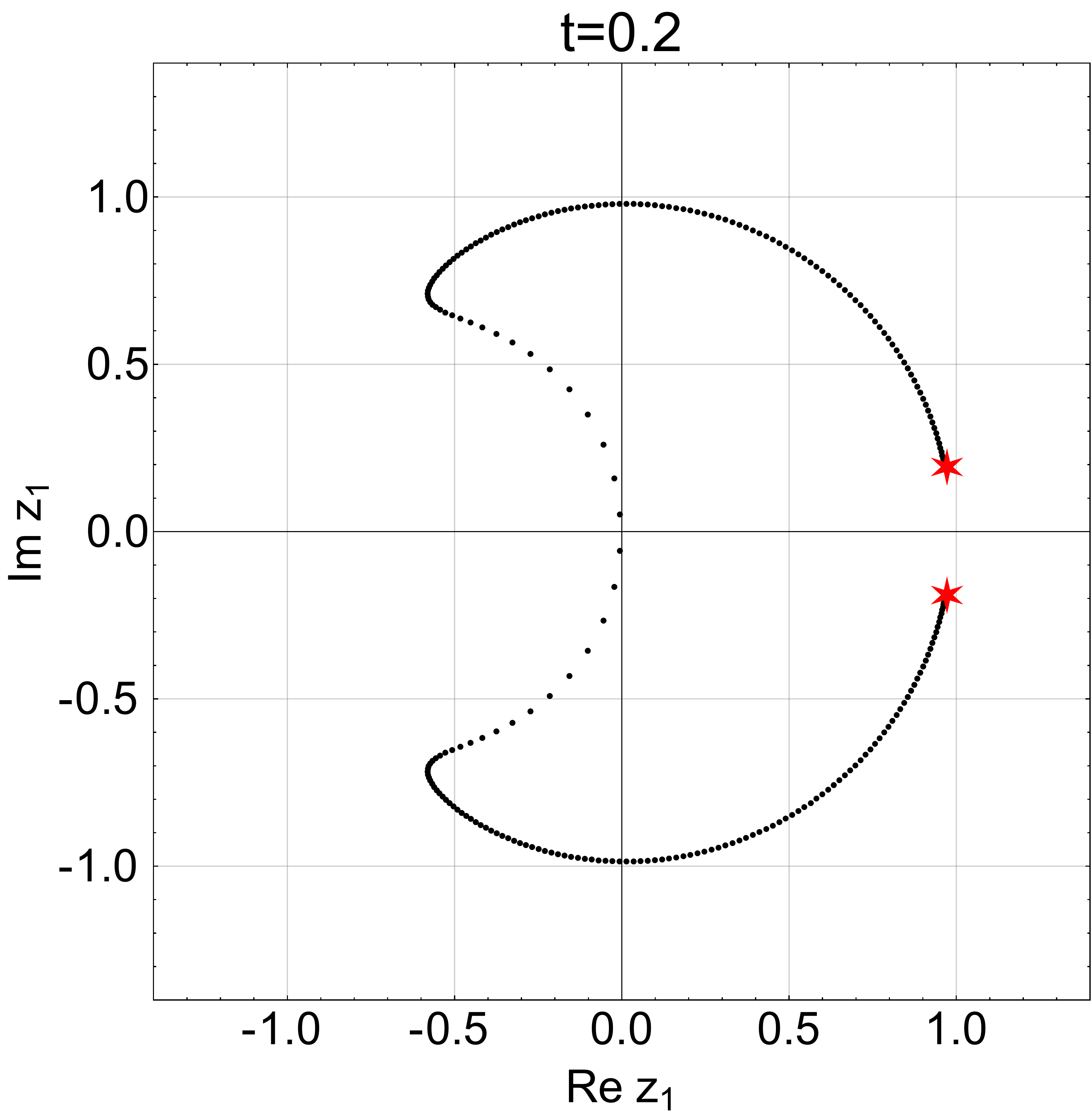}\\
		\hspace*{0.04\textwidth}(a)\hspace*{0.39\textwidth}(b)\\
		\vspace*{2mm}
		\includegraphics[width=0.35\columnwidth]{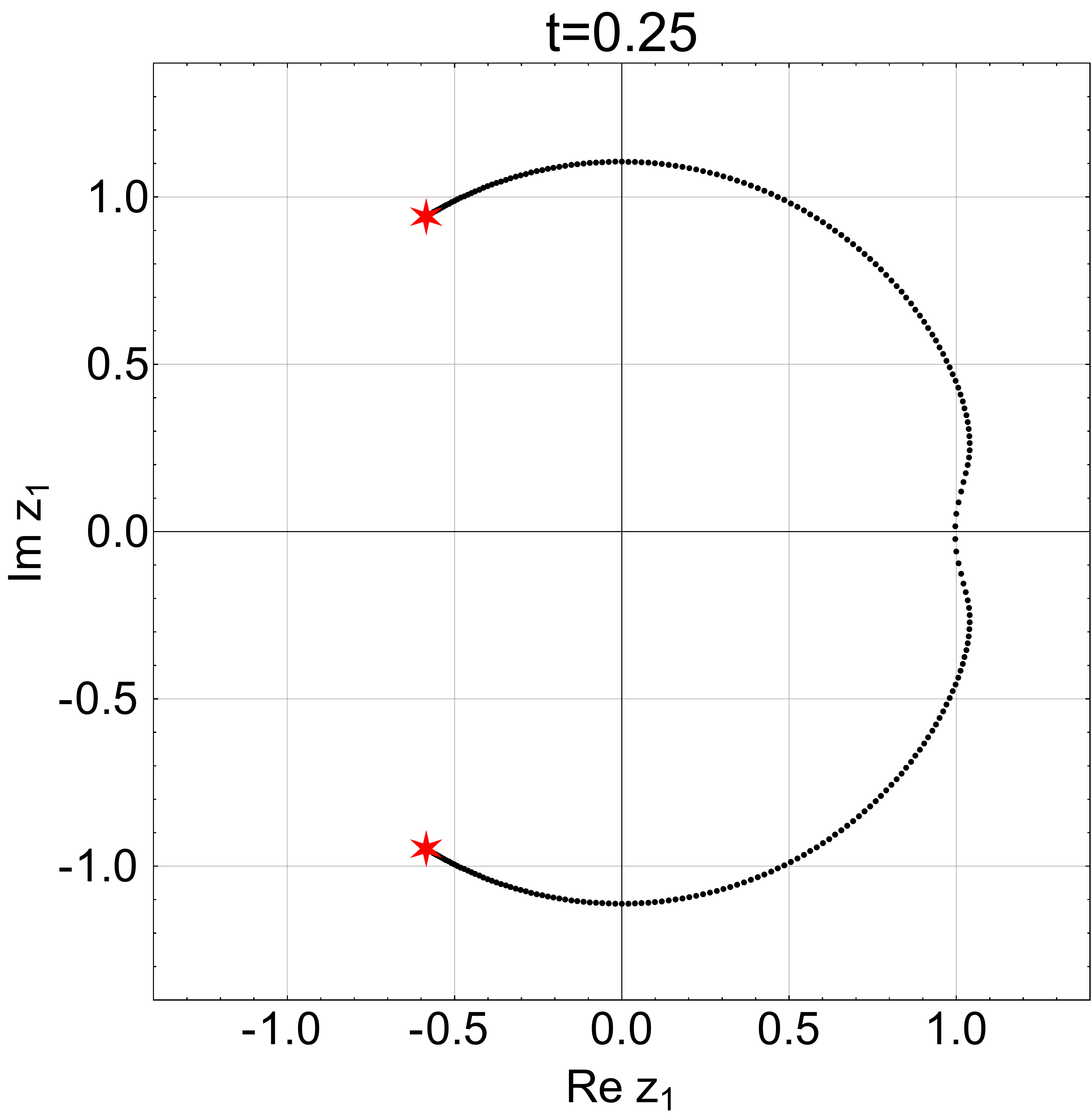}
		\includegraphics[width=0.35\columnwidth]{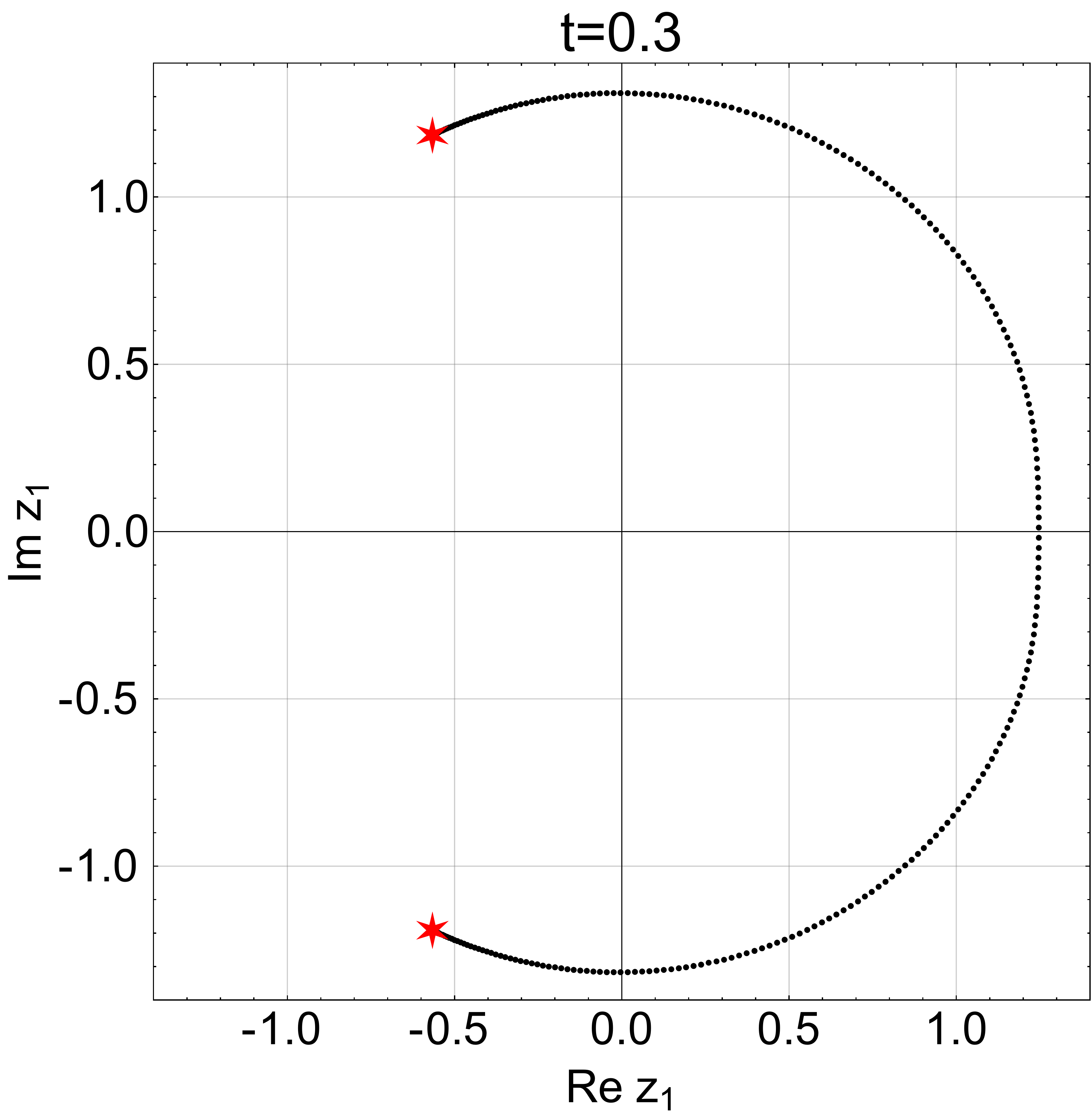}\\
		\hspace*{0.04\textwidth}(c)\hspace*{0.39\textwidth}(d)
		\caption{Lee-Yang zeros of the $(2,-5)$-state Potts model at different values of temperature (a)~$t=0.15$, (b)~$t=0.2$, (c)~$t=0.25$, (d)~$t=0.3$ in the complex $z_1-$plane for the system size $N=256$ \cite{Sarkanych18}. Panels (a) and (b) illustrate zeros below the zero-field critical temperature $t_c=0.25$, panel (c) and (d) illustrate zeros at and above $t_c$. Large red stars show edges and black dots ordinary Lee-Yang zeros. 
		\label{LYminusr}}
	\end{center}
\end{figure}

In the transfer-matrix approach, the coordinates of the partition function zeros are found from the 
condition that (at least) two largest eigenvalues of the transfer matrix are equal in absolute value. 
The partition function zeros were analysed in the complex-temperature plane in the case when
$h_2 = 0$. Since the famous Ising result for a partition function of a classical spin chain,
there exist a vast literature in the statistical mechanics of critical phenomena, including a number 
of famous theorems \cite{Landau} on why it is impossible to have a phase transition in one-dimensional classical systems 
with short-range interactions.
Usually phase transitions arise through a competition between entropy and energy.
It is now well known that the reason behind the no-go theorems is that there is an excess of entropy 
in 1D systems - entropy always wins over energy so that the delicate balance that gives a phase transition is never achieved.
 The results of Refs. \cite{Sarkanych17,Sarkanych18} show a way around the no-go theorems and 
show how one can induce a phase transition on 1D systems. 
Specifically, it has been shown that (i) a positive number of invisible states do not change 
the phase transition; (ii)
a negative number of invisible states induces a phase transition at a positive temperature;
(iii) a complex external field acting on invisible states has the same effect. Although
neither of these is directly physically accessible, but
both can be mapped to 1D quantum models in which phase transitions are physical. Below we provide
some illustrations for the above mentioned results. 
For convenience the next variables  are used:
\begin{equation}\label{7a}
t=e^{-\beta}\, , \, z_1=e^{\beta h_1}\, , \, z_2=e^{\beta h_2}\, .
\end{equation}

In Fig.~\ref{LYminusr} the Lee-Yang zeros in complex $z_1-$plane for the $(2,-5)$ ISPM for different values of temperature are shown.  
Loci of zeros cross the real axis at positive temperatures for
negative values of $r$. 
For small temperatures edge is located in the positive semi plane ${\rm Re}\,z_1>0$, while at  higher temperatures it jumps to the region ${\rm Re}\,z_1<0$. 
This jump occurs below the critical temperature $t_c$. 
The correlation length is infinite at the $t_c$ but the entropy has discontinuity. Therefore the phase transition can be interpret as of first order.

\begin{figure}[h]
	\begin{center}
		\includegraphics[width=0.6\columnwidth]{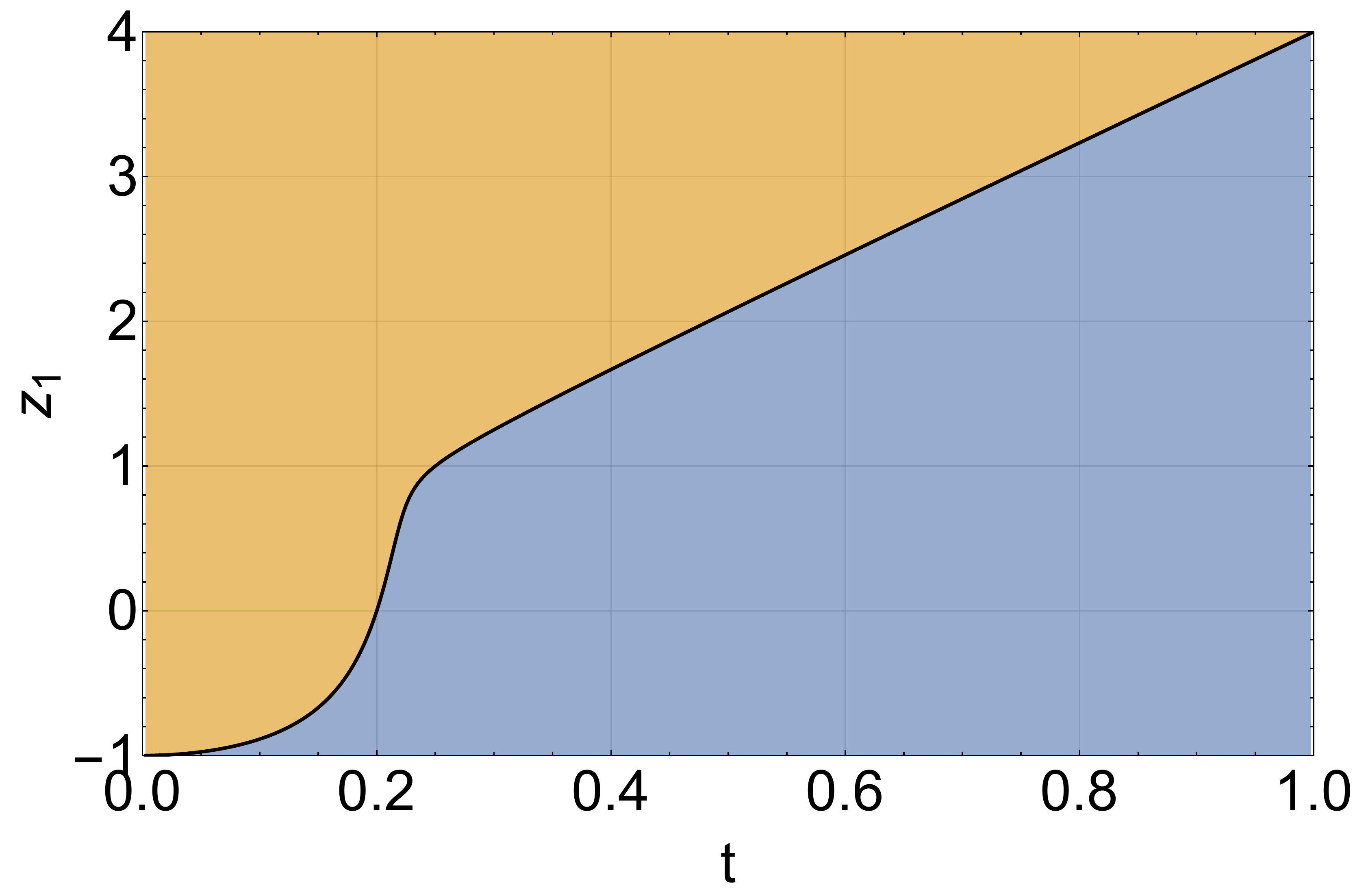}
		\caption{Phase diagram of the $(2,-5)$-state Potts model in $(t,z_1)$ plane  divided into 2 regions according to the maximal eigenvalue. Values $z_1=e^{\beta h_1}<0$ correspond to the complex values of magnetic field $h_1$, while $0<z_1<1$ corresponds to negative values of physical field $h_1$  \cite{Sarkanych18}. 
			\label{phasediagram}}
	\end{center}
\end{figure}

The set of crossing points for various temperatures in the range $0\leq t \leq 1$ can be interpreted as a phase diagram and is shown for the $(2,-5)$-state model as the solid black line in Fig. \ref{phasediagram}. 
The spontaneous transition is identified at $t=0.25$, $z_1=1$.
The counterpart for the ordinary Ising model is at  $t=0$, $z_1=1$ --- i.e., at vanishing instead of positive temperature.  
 
\begin{figure}[h]
	\begin{center}
		\includegraphics[width=0.5\columnwidth]{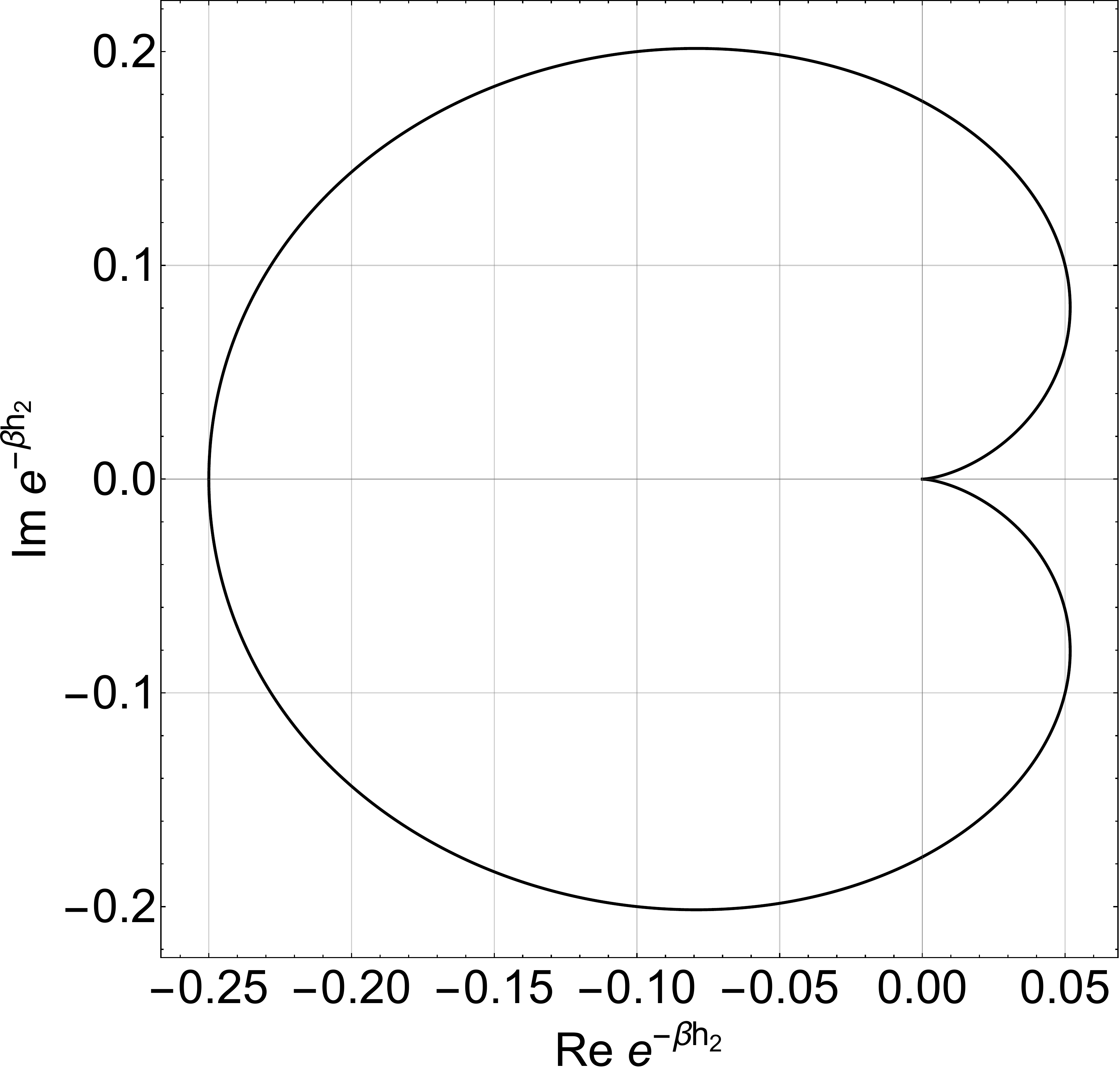}
		\caption{Values of the $e^{-\beta h_2}$ for which a phase transition occurs at positive temperature for the $(2,3)$ ISPM. Each point corresponds to a certain critical temperature value \cite{Sarkanych18}. 
			\label{complexh2}}
	\end{center}
\end{figure}

Another possibility to obtain a positive temperature phase transition is to consider complex external fields.
In Fig. \ref{complexh2} complex values of $e^{-\beta h_2}$ that can trigger a positive temperature phase transition in the $(2,3)$ ISPM are shown. 
Each point on the plot corresponds to a certain value of the critical temperature, and thus external magnetic field $h_2$.
 
\section{ISPM on a complete graph and on a scale-free network}\label{IV} 

In this section, we review the results of Refs. \cite{Krasnytska16,Sarkanych19,Sarkanych23}. There,
the ISPM was analysed within the mean field approach, having in mind two  different background 
structures where the interacting spins are located: the complete graph \cite{Krasnytska16} and the
scale-free network \cite{Sarkanych19,Sarkanych23}.

In the case of the complete graph topology, within mean-field approximation,
the ordinary Potts model is characterised by second order phase transition for $q\leq2$ and first order transition 
when $q>2$. Since adding invisible states makes the phase transition only sharper, the goal of Ref. \cite{Krasnytska16} 
was to pay special attention to the region $q\leq 2$ where the second order transition occurs. For  
$q=2$  Potts model the result of Ref. \cite{Ananikian13} was recovered: 
the phase transition changes from the first to the second order when the number of invisible states equals $r_c=3.62$.
\begin{figure}[h!]
\centerline{\includegraphics[width=0.5\columnwidth]{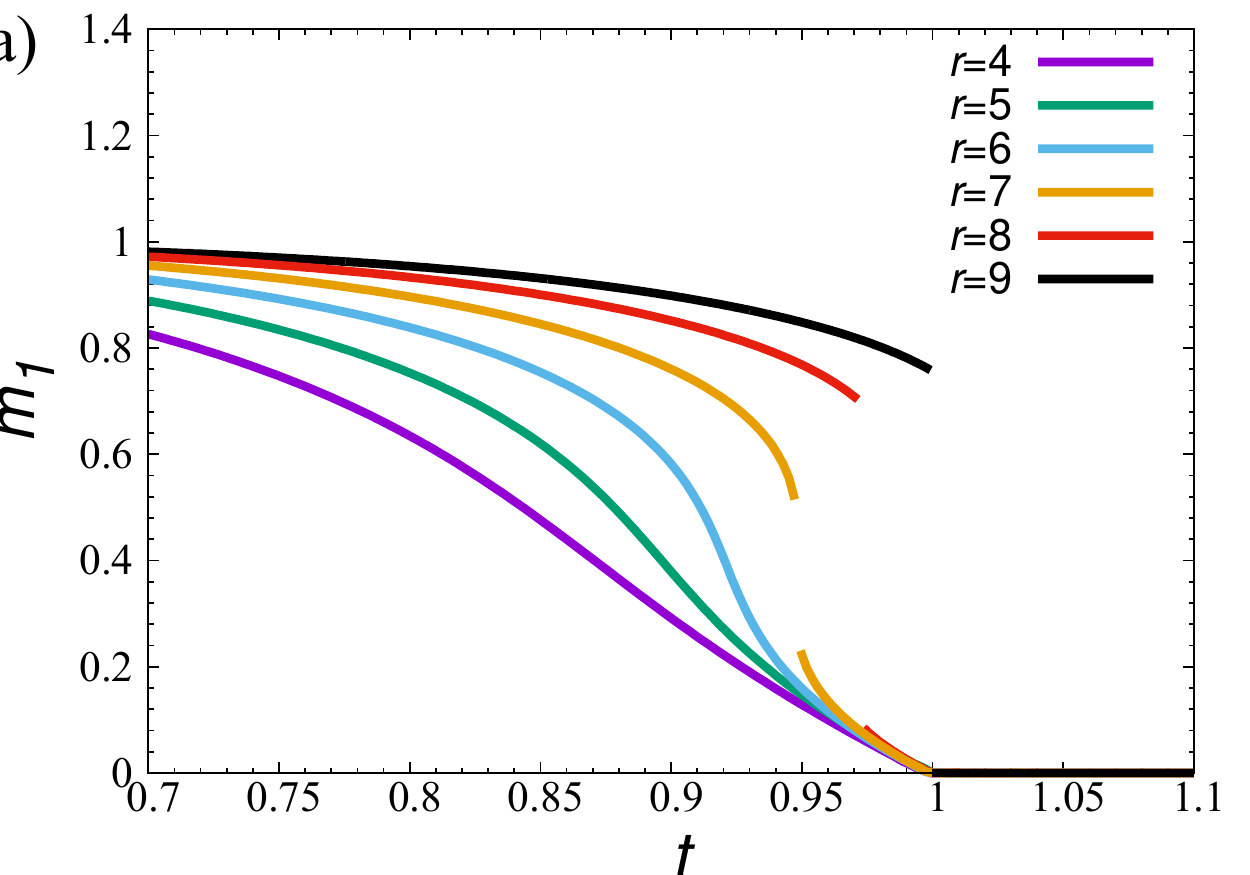} \includegraphics[width=0.5\columnwidth]{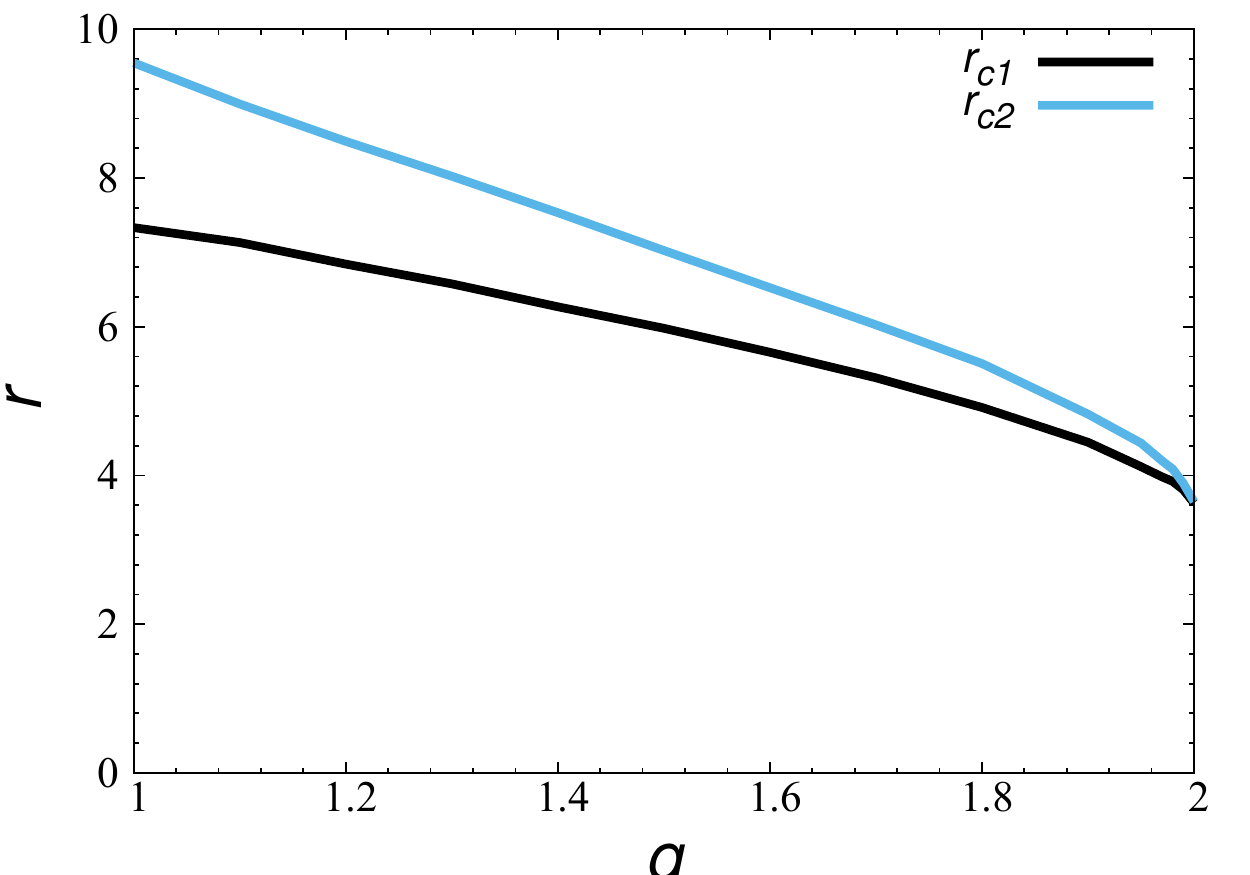}}
\centerline{ (a) \hspace{3cm} (b)}
    \caption{ (a) Dependencies of the order parameter  $m_1$ on the reduced temperature $t$ for $r=4,5,6,7,8,9$ for $q=1.2$ \cite{Krasnytska16}. 
    (b) The phase diagram of the ISPM on the complete graph in the $(r,q)$ plane. Transition is of
the first order for $r>r_{c2}$ and of the second order for $r<r_{c1}$. Both first and second order phase transitions occur 
in the intermediate region  $r_{c1}<r<r_{c2}$  \cite{Krasnytska16}.}
    \label{4a}
\end{figure}
The typical behaviour of the order parameter at different $r$ at $1<q<2$ is shown in Fig.\ref{4a} (a). There, we display the
temperature behaviour of the order parameter $m_1(t)$, $t=(T_c-T)/T_c$, 
that  emerges below the transition point and breaks the
system symmetry at $q=1.2$. As one can see from the figure, there are two marginal dimensions $r_{c1}$ and $r_{c2}$ 
that separate regions with different critical  behaviours.
For low $r$, the system undergoes only second order phase transitions.
If $r$ is large on the other hand there is only a first order phase transition.
Between these two regimes, the system is characterised by both a first and a second order phase transition occurring at different temperatures.
Finding those marginal values in the range $1<q<2$ resulted in the phase diagram shown in Fig.\ref{4a} (b) \cite{Krasnytska16}.  
Two conclusions can be made from the phase diagram. 
First, in the limit $q=2$ the two marginal dimensions coincide, $r_{c1}=r_{c2}$. 
Thus in this particular case the transition changes from the second  to the first order.
The second conclusion concerns the limit  $q\to 1$. This limit relates Potts model to the problem of 
bond percolation. Therefore, the results obtained in Ref. \cite{Krasnytska16} suggest that
adding invisible states may turn the transition in the bond-percolation model into a strong 
first-order regime. Currently, several mechanisms that sharpen the percolation transition are 
discussed, see e.g. \cite{Achlioptas09,Bastas14,Adler91,Buldyrev10,Lee16,Lee17,Herrmann16}. 
The scenario based on the ISPM adds one more mechanism to this picture.

Refs. \cite{Sarkanych19,Sarkanych23} deal with an ISMP for which the interacting spins are located on the nodes of a complex network.
Studies of phase transitions on complex networks attracted much attention for different reasons
\cite{Dorogovtsev08}, being important in particular to analyse opinion formation or percolation-related processes 
on networks (network resistance, epidemic spreading, etc). The Hamiltonian (\ref{1.1}) of the  $(q,r)$ ISPM can be rewritten
for an arbitrary graph  as \cite{Krasnytska16}:
\begin{equation}\label{1a}
 H= - \sum_{<i,j>}J _{ij}\sum_{\alpha=1}^q \delta
_{S_i,\alpha}\delta _{\alpha,S_j} \, , \hspace{1cm} S_i=(1,...,q,q+ 1,...,q+r).
\end{equation}
Here, $J_{ij}$ are the elements of the graph adjacency matrix: they  are equal to $1$ when  there is a link between nodes $i$ and $j$ and $0$ otherwise. 
Considering the probability for two nodes to be connected as $p_{ij}$ one defines:
  \begin{eqnarray}\label{11a}
J_{ij} =\left\{
\begin{array}{cll}
                & 1\, ,   & p_{ij}, \\
                & 0\, , & 1-p_{ij}.
              \end{array}
  \right.
\end{eqnarray}
For an annealed network the probability for two nodes to be connected is linearly proportional to their degrees (number of nearest neighbours) \cite{Lee09}: 
\begin{equation}
    p_{ij}=k_i k_j /(N\bar{k}),
\end{equation}
where $k_i$ and $\bar{k}$ are respectively the degree of node $i$ and the average node degree. 

The ISPM  is described by two global order parameters: $m_1$ and $m_2$.  For the inhomogeneous
network they are introduced as  linear combinations of the local order parameters $m_{1i}$ and 
$m_{2i}$ defined on each of the nodes 
with weights proportional to the node degree (see \cite{Sarkanych19} for more details):
\begin{equation}\label{11b}
m_1=\frac{\sum_i k_i m_{1i}}{\sum_i k_i}, \hspace{2cm}
m_2=\frac{\sum_i k_i m_{2i}}{\sum_i k_i}.
\end{equation}

The free energy density  in terms of the order parameters reads \cite{Sarkanych19}: 
\begin{eqnarray} \nonumber
    f(m_1,m_2)&=&\frac{\bar{k}}{(q+r)^2}\Big((rm_2+1+(q-1)m_1)^2+ (q-1)(rm_2+1-     \\ && \nonumber
    (r+1)m_1)^2\Big)-\frac{1}{\beta}\int_{k_{\rm min}}^\infty dk P(k)
\ln\Big(e^{\beta(\frac{k}{q+r}(m_1(q-1)+1+rm_2))}+
  \\ &&  \label{ff0}  (q-1)e^{\frac{\beta
  k}{q+r}(m_2r+1-(r+1)m_1)}+r\Big),  
\end{eqnarray}
where the integration is performed over all node degrees $k_{\rm min}<k<\infty$ and $P(k)$ is the node degree distribution. 
Of particular interest are scale-free networks \cite{Albert02,Goltsev03,Dorogovtsev08}, when the node-degree distribution decay is governed by a
power law:
\begin{equation}
      P(k)\propto k^{-\lambda}\, , \, k \to \infty.
        \end{equation}
The resulting expression for the free energy depends on two global order parameters $m_1$ and $m_2$, that describe the state of the system and the model parameters, such as the numbers of states $q$ and $r$, the temperature $\beta$ and the decay exponent $\lambda$ describing the topology of the system. 
It is well established by now, that when the critical behaviour on  scale-free networks is considered, the decay exponent 
 $\lambda$ is a global parameter playing a role similar to space dimension for lattices \cite{Igloi02,Dorogovtsev04,Kasteleyn69}. 
In order to find what is the phase diagram one has to minimize the free energy with respect to the order parameters.
Fixing the values of $q$, $r$ and $\lambda$ one can sweep through a certain region of temperatures to calculate the values 
of $m_1$ and $m_2$ that minimize the free energy. 
In this way, the behaviour of the order parameters vs temperature allows to obtain the order of the phase transition.
In the limit $r\to 0$ the ISPM reduces to the standard Potts model and its behaviour on a scale free network is known \cite{Igloi02,Krasnytska13}.
It is determined by $q$ and $\lambda$: 1)  for $2<\lambda<3$ for any $q$ the system is ordered at any finite temperature and 
no phase transition is observed; 2) for $\lambda>3$ and $1\leq q\leq2$ the second order phase transition occurs; 3) for $\lambda>3$ and $q>2$ either the first or the second order phase transition occur and $\lambda_c(q)$ separates those regions. 
    \begin{figure}[h!]
\centerline{\includegraphics[width=0.6\columnwidth]{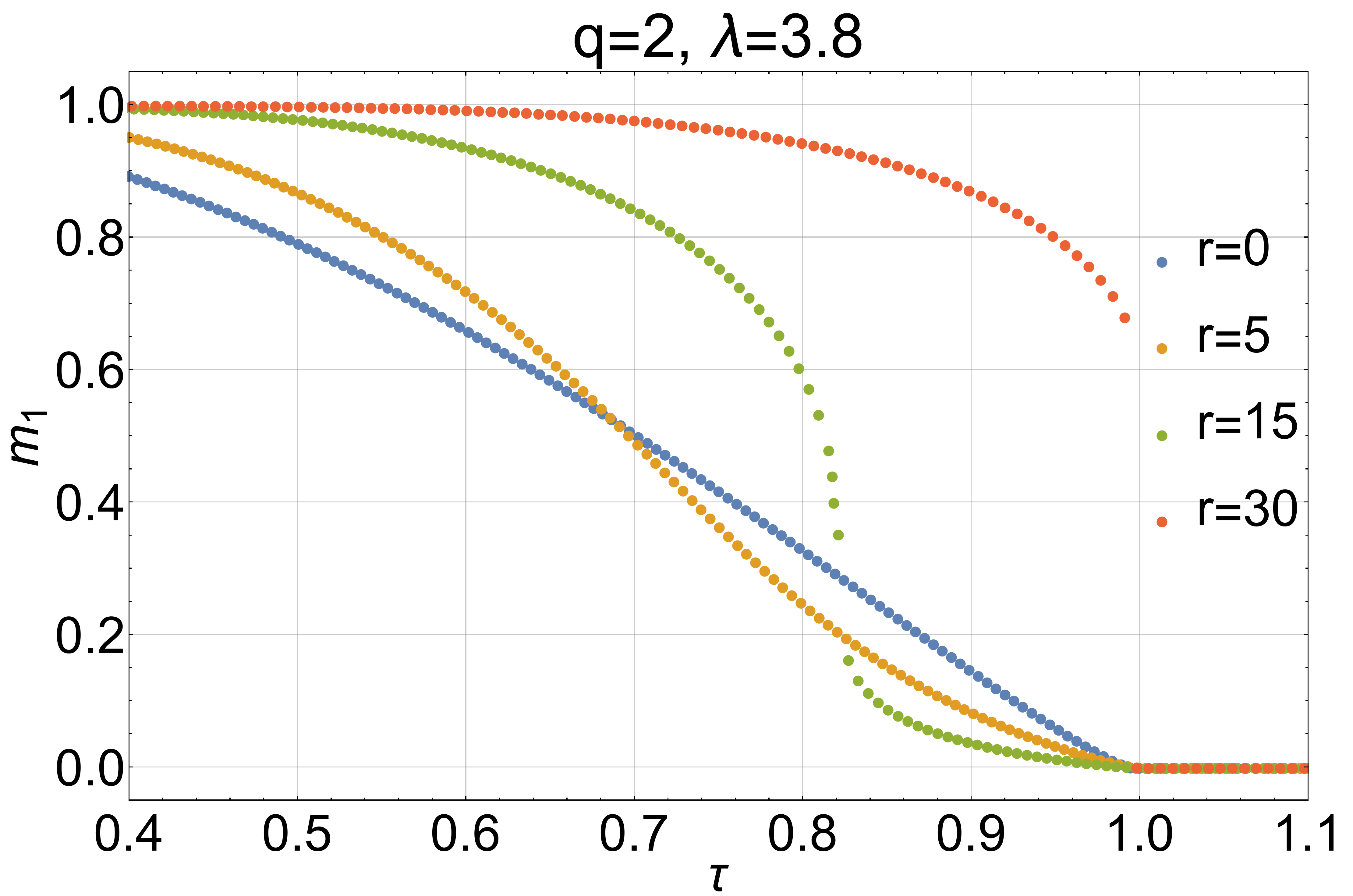}}
    \caption{Order parameter $m_1$ as a function of the reduced temperature $\tau=T/T_c$ for various values of $r$ and fixed  $q=2$, $\lambda=3.8$.
        Different values of $r$ lead to different critical regimes \cite{Sarkanych19}.}
    \label{m1t}
\end{figure}  
In the case of the ISPM on a scale free network there are three parameters $q$, $r$ and $\lambda$ that define the critical behaviour.
The typical behaviour of the order parameter is given in Fig. \ref{m1t}. 
In the region $1<q\leq2$ at $\lambda>3$ one observes that adding the invisible states strengthens ordering and leads to a first order phase transition. 

 \begin{figure}[h!]
\centerline{ \includegraphics[width=0.5\columnwidth]{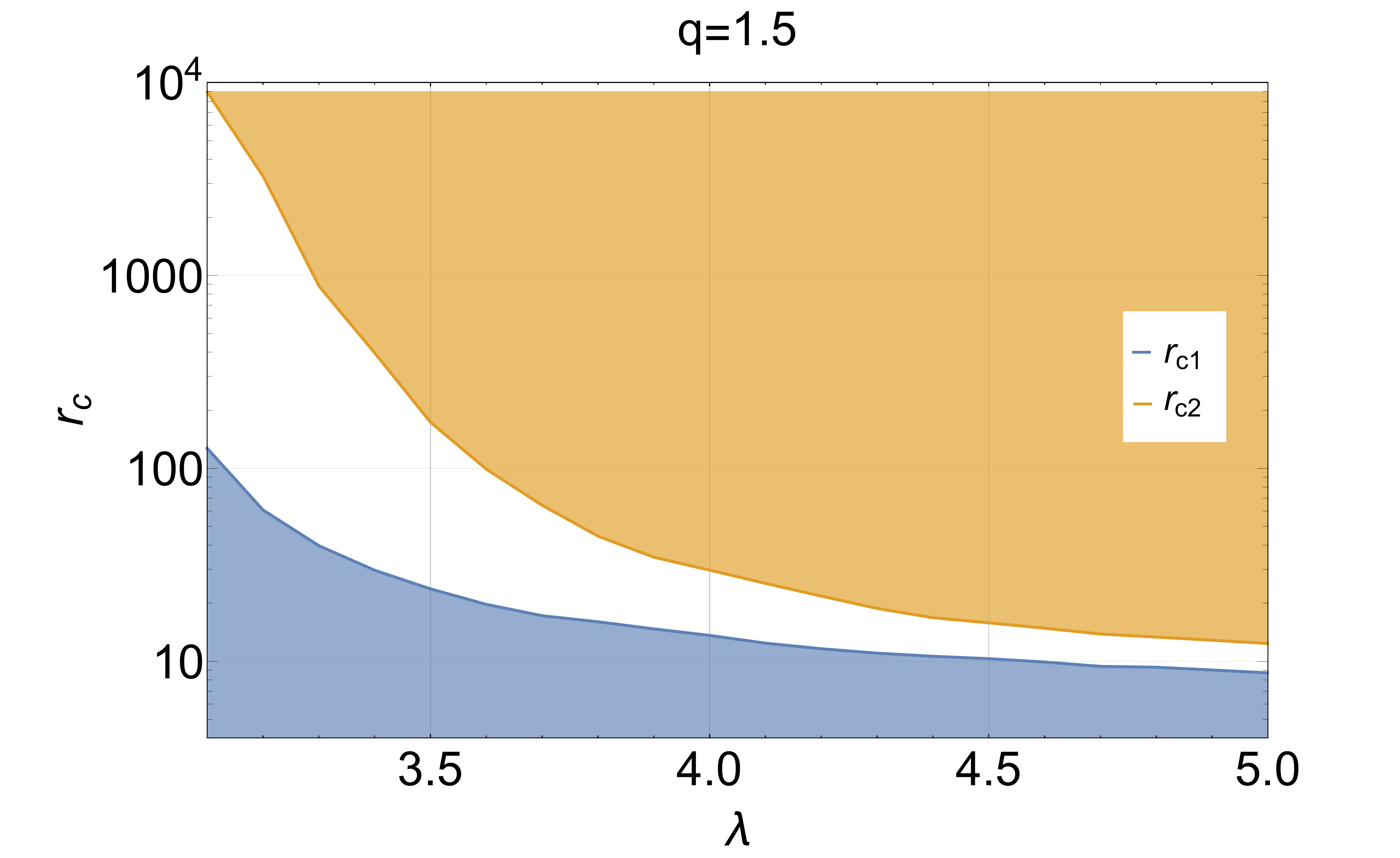}\includegraphics[width=0.5\columnwidth]{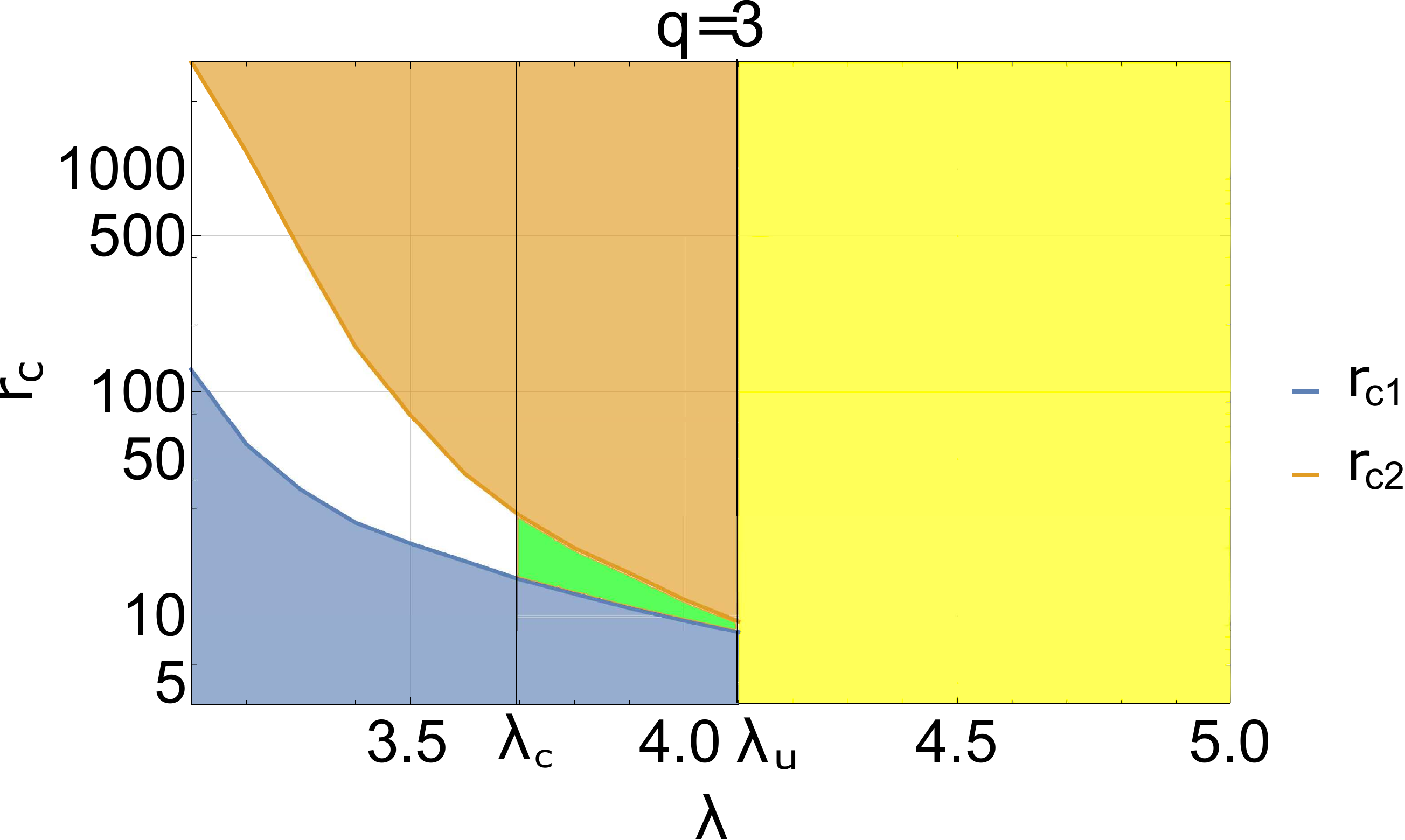}}
\centerline{ (a) \hspace{3cm} (b)}
    \caption{The phase diagram of the ISPM on a scale-free network for (a) $q=1.5$ \cite{Krasnytska22},  (b) $q=3$ \cite{Krasnytska22}.  
    (a) In  the  lower  (blue)  region  a  system  possesses  only  the  second  order phase transition; in the region in-between 
    the lines, there are both first and second order phase transitions at different temperatures; in the upper region (yellow) 
    only first order phase transition occurs. (b) The phase diagram has two marginal values $\lambda_c$ and $\lambda_u$. 
    For $\lambda<\lambda_c$ the behaviour is identical to (a). For $\lambda>\lambda_u$ there is only a first order phase transition. 
    In the region in between $\lambda_c\leq\lambda\leq\lambda_u$ for low and high values of $r$ only first order phase transition 
    is observed, while for $r_{c1}\leq r\leq r_{c2}$ two first order phase transitions occur. } 
    \label{phasediag}
\end{figure}  

The phase diagram of the ISPM on a scale-free network is shown in Fig. \ref{phasediag}. 
It is similar to that for the complete graph shown in Fig.  \ref{4a}(a) \cite{Krasnytska16}. 
In particular, as one can see  from Fig. \ref{phasediag}, there are two sets of marginal 
values $r_{c1} (\lambda)$ and $r_{c2} (\lambda)$ that separate  regions where only a second 
or a first order phase transition is observed and a region where they co-exist.
One can conclude that the smaller the values of $q$ or $\lambda$ the larger the number of invisible states 
is required to change the order of the phase transition.  

The case $q>2$ appears to be is even more involved \cite{Sarkanych23}. 
In this region, besides the two marginal values $r_{c1}$ and $r_{c2}$, there are two marginal values $\lambda_c$ and $\lambda_u$. 
Above $\lambda_u$ only a first order phase transition occurs. 
Below $\lambda_c$ the behaviour is qualitatively the same as in $1<q\leq 2$.
In between these two values $\lambda_c\leq\lambda\leq\lambda_u$ there are also three regions: 
for $r<r_{c1}$ and $r>r_{c2}$ there is only a first order phase transition, while for $r_{c1}\leq r\leq r_{c2}$ there are two first order phase transitions occurring at different temperatures \cite{Krasnytska22}. 
This hints that the topological influence, presented by the marginal value of  $\lambda_c(q)$ and $\lambda_u(q)$ is to a certain extent independent from an entropic one, presented by the number of invisible states $r$.
 
\section{Conclusions and outlook}\label{V}

Since its introduction \cite{Tamura2010,Tanaka2011} the ISPM  has attracted significant attention.
It was considered on different geometries from simple 1D chain \cite{Sarkanych17,Sarkanych18} to complex networks \cite{Sarkanych19,Krasnytska22} and using various methods from rigorous proofs \cite{Enter11,Enter12,Iakobelli12,Mori12} and exact results \cite{Sarkanych17,Sarkanych18} to Monte-Carlo simulations \cite{Tamura2010,Tanaka11a}. 
It is partially because the model allows for a clear way to regulate the entropy of the system through the number of invisible states.
Even a negative number of invisible states was considered to simulate the reduction of the entropy \cite{Sarkanych17,Sarkanych18}.
Due to the possibility to influence the energy-entropy interplay, the ISPM allows to tune the phase transition picture: 
addition of invisible states can make the phase transition sharper or even change a second order transition into a first order one.

Recently, the ISPM has been used to interpret the possible changes in the order of the percolation transition.
Although the percolation transition is known to be a continuous one \cite{Stauffer_book}, in the real world there
exist connectivity problems that are clearly abrupt and need conceptual understanding and quantitative treatment 
\cite{Herrmann16}.
The results obtained for ISPM on the complete graph \cite{Krasnytska16} prove that the addition of invisible states can lead a 
phase transition in the percolation limit $q\to 1$ of the standard Potts model to become a first order transition.
In this way one more mechanism is suggested to achieve a discontinuous percolation transition. This  mechanism differs
from those of explosive percolation \cite{Achlioptas09,Bastas14}, bootstrap percolation \cite{Adler91}, cascade 
of failures in interdependent networks  \cite{Buldyrev10} and hybrid percolation on multiplex interdependent Erd\H{o}s-R\'enyi networks \cite{Lee16,Lee17}.

Obviously, the idea to tune the  energy-entropy interplay by introducing novel
models appeared also before the ISMP was suggested. An example is given by 
 Ref. \cite{Ananikyan90}, where  a model for helix-coil transition in polypeptide chain was proposed.
It can be interpreted as the ISPM with multi-spin interaction and only a single visible state.  
Since polypeptides can be seen as  one-dimensional objects, the transfer matrix approach was used to obtain exact results.
Later this model was extended in Refs. \cite{hairyan1995,morozov2004,morozov2005,badasyan2005,grigoryan2007,Badasyan10,Badasyan11,Badasyan12}.
Moreover, similar ideas continue to be the subject of analysis. A recent example is given by 
the randomly colored Potts model \cite{Schreiber22}.
In this model, a random concentration $p$ of Potts spins
assumes $q_0$ states while the remaining $1-p$ of spins assume $q > q_0$
states. It is argued that changes in concentration may influence the phase
transition.

The invisible states, as  they are formulated in statistical physics, have their counterparts in other fields. In game theory, an analogy of invisible states is given by the `neutral strategies'. The latter are such strategy choices that provide zero payoffs regardless of the 
player's move, see \cite{Kiraly19} and references therein. Another example is `invisible fluctuations' that has been introduced \cite{Tanaka13DiversQuant} to solve optimization problems by annealing methods \cite{Kurihara09,Sato09,Tamura14,Tanaka13DiversQuant,Tanaka13}.

\bmhead{Aknowledgements}
It is our pleasure and honour for us to contribute by this paper to the Festschrift dedicated to Malte
Henkel on the occasion of his jubilee. Doing so we acknowledge Malte's contributions to many problems in
the field of phase transitions and criticality mentioned (and not mentioned) in this review \cite{Helkel1,Helkel2}.
MK thanks the National Academy of Sciences of Ukraine,  grant for research laboratories/groups of 
young scientists No 07/01-2022(4); the PAUSE program and hospitality at LPCT, Lorraine University.

\bmhead{Data Availability Statement} No Data associated in the manuscript.
 
\bibliography{sn-bibliography}

\end{document}